\documentclass[11pt]{article}

\usepackage{graphicx}
\usepackage{amsfonts}

\usepackage{theorem}
\newtheorem{Lem}{Lemma}[section]

\newtheorem{The}[Lem]{Theorem}

\newtheorem{Cor}[Lem]{Corollary}

\newtheorem{Rem}[Lem]{Remark}

\newcommand{\qed}{\hbox{\rule{6pt}{6pt}}}

\begin{document}
\title{Matrix trace inequalities on the Tsallis entropies}

\author{Shigeru Furuichi$^1$\footnote{E-mail:furuichi@chs.nihon-u.ac.jp}\\
$^1${\small Department of Computer Science and System Analysis, 
College of Humanities and Sciences,}\\{\small Nihon University,
3-25-40, Sakurajyousui, Setagaya-ku, Tokyo, 156-8550, Japan}}


\date{}
\maketitle
{\bf Abstract.} 
Maximum entropy principles in nonextensive statistical physics are revisited as an application of the Tsallis relative entropy defined for
non-negative matrices in the framework of matrix analysis. 
In addtition, some matrix trace inequalities related to the Tsallis relative entropy are studied.
\vspace{3mm}


\vspace{3mm}


\vspace{3mm}

\begin{center}
{\it Dedicated to Professor Kunio Oshima on his 60th birthday}\\
\end{center}
\section{Introduction}
In 1988, Tsallis introduced the one-parameter extended entropy for the analysis of a physical model in statistical physics \cite{Tsa}.
In our previous papers, we studied the properties of the Tsallis relative entropy \cite{FYK,Furu} and the Tsallis relative operator entropy \cite{YKF,FYK2}. 
The problems on the maximum entropy principle in Tsallis statistics have been studied for classical systems and quantum systems \cite{MNPP,TMP,AMPP,Abe}. 
Such problems were solved by the use of the Lagrange multipliers formalism. We give a new approach to such problems, 
that is, we solve them by applying the non-negativity of the Tsallis relative entropy without using the Lagrange multipliers formalism.
In addition, we show further results on the Tsallis relative entropy.

In the present paper, the set of $n \times n$ complex matrices is denoted by $M_n(\mathbb{C})$. 
That is, we deal with $n \times n$ matrices because of Lemma \ref{n_lemma} in section 2. However some results derived in the present paper also hold
for the infinite dimensional case.
In the sequel, the set of all density matrices (quantum states) is represented by 
$$D_n(\mathbb{C}) \equiv \left\{ X \in M_n(\mathbb{C}) :X \geq 0, Tr[X] = 1 \right\}.$$
$X \in M_n(\mathbb{C})$ is called by a non-negative matrix and denoted by
$X \geq 0$, if we have $\langle X x,x\rangle \geq 0$ for all $x \in \mathbb{C}^n$. 
That is, for a Hermitian matrix $X$, $X \geq 0$ means that all eigenvalues of $X$ are non-negative.
In addition, $X \geq Y $ is defined by $X-Y \geq 0$.
For $-I \leq X \leq I$ and $\lambda \in (-1,0) \cup (0,1)$, we denote the generalized exponential function by
 $\exp_{\lambda}\left(X\right) \equiv \left(I+\lambda X\right)^{1/{\lambda}}$. As the inverse function of 
$\exp_{\lambda}(\cdot)$, for $X \geq 0$ and $\lambda \in (-1,0) \cup (0,1)$,  
we denote the generalized logarithmic function by $\ln_{\lambda} X \equiv \frac{X^{\lambda}-I}{\lambda}$.
Then the Tsallis relative entropy and the Tsallis entropy for non-negative matrices $X$ and $Y$ are defined by
$$
D_{\lambda}(X\vert Y) \equiv Tr \left[X^{1-\lambda} \left( \ln_{\lambda}X -\ln_{\lambda}Y  \right) \right], \quad S_{\lambda}(X) \equiv - D_{\lambda}(X\vert I).
$$
These entropies are generalizations of the von Neumann entropy \cite{von} and of the Umegaki relative entropy \cite{Ume} in the sense that
$$\lim_{\lambda\to 0}S_{\lambda}(X) = S_0(X) \equiv -Tr[X \log X]$$
and 
$$\lim_{\lambda\to 0}D_{\lambda}(X\vert Y) = D_0(X\vert Y) \equiv Tr[X (\log X-\log Y)].$$

\section{Maximum entropy principle in nonextensive statistical physics}
In this section, we study the maximization problem of the Tsallis entropy with the constraint on the $\lambda$-expectation value.
In quantum system, the expectation value of an observable (a Hermitian matrix) $H$ in a quantum state (a density matrix) $X \in D_n(\mathbb{C})$
is written as $Tr[XH]$. Here, we consider the $\lambda$-expectation value $Tr[X^{1-\lambda}H]$ as a generalization of the usual expectation value.
Firstly, we impose the following constraint on the maximization problem of the Tsallis entropy:
$$\widetilde{ C_{\lambda} } \equiv \left\{ X \in D_n(\mathbb{C}) : Tr[X^{1-\lambda}H] = 0 \right\},$$
for a given $n\times n$ Hermitian matrix $H$.
We denote a usual matrix norm by $\left\|\cdot \right\|$, namely for $A\in M_n(\mathbb{C}) $ and $x \in \mathbb{C}^n$,   
$$\left\|A \right\| \equiv \max_{\left\|x\right\|=1}\left\|Ax\right\|.$$ 
Then we have the following theorem.
\begin{The}   \label{the2_1}
Let $Y= Z_{\lambda}^{-1} \exp_{\lambda}\left(-H/{\left\| H\right\|} \right) $, 
where $Z_{\lambda} \equiv Tr[\exp_{\lambda}\left(-H/{\left\| H\right\|} \right) ]$,
 for an $n\times n$ Hermitian matrix $H$ and $\lambda \in (-1,0) \cup (0,1)$.
If $X \in \widetilde{ C_{\lambda} }$, then $S_{\lambda}(X) \leq - c_{\lambda}\ln_{\lambda}Z_{\lambda}^{-1},$ where $c_{\lambda} \equiv Tr[X^{1-\lambda}].$
\end{The}
{\it Prrof}:
Since $Z_{\lambda} \geq 0$ and we have $\ln_{\lambda}(x^{-1}Y) = \ln_{\lambda}Y + (\ln_{\lambda}x^{-1})Y^{\lambda}$
 for a non-negative matrix $Y$ and scalar $x$,
we calculate
\begin{eqnarray*}
Tr[X^{1-\lambda}\ln_{\lambda}Y] &=& Tr[X^{1-\lambda}\ln_{\lambda}\left\{Z_{\lambda}^{-1}\exp_{\lambda}\left(-H/\left\|H\right\|\right) \right\}]\\
&=& Tr[X^{1-\lambda}\left\{ -H/\left\| H\right\| +\ln_{\lambda}Z_{\lambda}^{-1} \left(I-\lambda H/\left\|H\right\|\right) \right\}]\\
&=& Tr[X^{1-\lambda}\left\{ \ln_{\lambda}Z_{\lambda}^{-1}I -Z_{\lambda}^{-\lambda} H/\left\|H \right\| \right\}]\\
&=& c_{\lambda}\ln_{\lambda}Z_{\lambda}^{-1},
\end{eqnarray*}
since $\ln_{\lambda}Z_{\lambda}^{-1} = \frac{Z_{\lambda}^{-\lambda}-1}{\lambda}$ by the definition of the generalized logarithmic function $\ln_{\lambda}(\cdot)$.
By the non-negativity of the Tsallis relative entropy:
\begin{equation} \label{nonnega_rela}
Tr[X^{1-\lambda}\ln_{\lambda}Y] \leq Tr[X^{1-\lambda}\ln_{\lambda}X],
\end{equation}
we have 
$$
S_{\lambda}(X) = -Tr[X^{1-\lambda}\ln_{\lambda}X] 
\leq -Tr[X^{1-\lambda}\ln_{\lambda}Y] 
= -c_{\lambda}\ln_{\lambda}Z_{\lambda}^{-1}.
$$
\hfill \qed

Next, we consider the slightly changed constraint:
$$C_{\lambda} \equiv \left\{ X \in D_n(\mathbb{C}): Tr[X^{1-\lambda}H] \leq 
Tr[Y^{1-\lambda}H]\,\,and\,\,Tr[X^{1-\lambda}]\leq Tr[Y^{1-\lambda}] \right\}$$
for a given $n\times n$ Hermitian matrix $H$, as the maximization problem for the Tsallis entropy.
 To this end, we prepare the following lemma. 

\begin{Lem} \label{n_lemma}
For a given $n \times n$ Hermitian matrix $H$, if $n$ is a sufficient large integer, then we have $Z_{\lambda} \geq 1$.
\end{Lem}
{\it Proof}:
\begin{itemize}
\item[(i)] For a fixed $0 < \lambda <1$ and a sufficient large $n$, we have 
\begin{equation} \label{n_lem_ineq1}
\left(1/n\right)^{\lambda} \leq 1 - \lambda.
\end{equation}
From the inequalities $  -\left\|  H \right\|I\leq H \leq \left\|  H \right\|I$, we have
\begin{equation} \label{n_lem_ineq3}
 (1-\lambda)^{\frac{1}{\lambda}}I \leq \exp_{\lambda}\left( -H/\left\|H\right\|\right) \leq  (1+\lambda)^{\frac{1}{\lambda}}I.
\end{equation}
By inequality (\ref{n_lem_ineq1}), we have
$$\frac{1}{n}I\leq  (1-\lambda)^{\frac{1}{\lambda}}I \leq \exp_{\lambda}\left( -H/\left\|H\right\|\right), $$
which implies $Z_{\lambda} \geq 1$.
\item[(ii)] For a fixed $-1 < \lambda <0$ and a sufficient large $n$, we have 
\begin{equation} \label{n_lem_ineq2}
\left(1/n\right)^{\lambda} \geq 1 - \lambda.
\end{equation}
Analogously to (i), we have inequalities (\ref{n_lem_ineq3}) for $-1 < \lambda <0$.
By inequality (\ref{n_lem_ineq2}), we have
$$\frac{1}{n}I\leq  (1-\lambda)^{\frac{1}{\lambda}}I \leq \exp_{\lambda}\left( -H/\left\|H\right\|\right), $$
which implies $Z_{\lambda} \geq 1$.
\end{itemize}
\hfill \qed

Then we have the following theorem by the use of Lemma \ref{n_lemma}.

\begin{The} \label{MEPthe1}
Let $Y= Z_{\lambda}^{-1} \exp_{\lambda}\left(-H/{\left\| H\right\|} \right) $, 
where $Z_{\lambda} \equiv Tr[\exp_{\lambda}\left(-H/{\left\| H\right\|} \right) ]$,
for $\lambda \in (-1,0) \cup (0,1)$ and an $n \times n$ Hermitian matrix $H$. 
If $X \in C_{\lambda} $ and $n$ is sufficient large, then $S_{\lambda}(X) \leq S_{\lambda}(Y)$.
\end{The}
{\it Proof}]
Due to Lemma \ref{n_lemma}, we have $\ln_{\lambda}Z_{\lambda}^{-1} \leq 0$ for a sufficient large $n$.
Thus we have $ \ln_{\lambda}Z_{\lambda}^{-1}Tr[X^{1-\lambda}] \geq \ln_{\lambda}Z_{\lambda}^{-1}Tr[Y^{1-\lambda}]$ for $X\in C_{\lambda}$.
As similar way to the proof of Theorem \ref{the2_1}, we have 
\begin{eqnarray*}
Tr[X^{1-\lambda}\ln_{\lambda}Y] &=& Tr[X^{1-\lambda}\ln_{\lambda}\left\{Z_{\lambda}^{-1}\exp_{\lambda}\left(-H/\left\|H\right\|\right) \right\}]\\
&=& Tr[X^{1-\lambda}\left\{ -H/\left\| H\right\| +\ln_{\lambda}Z_{\lambda}^{-1} \left(I-\lambda H/\left\|H\right\|\right) \right\}]\\
&=& Tr[X^{1-\lambda}\left\{ \ln_{\lambda}Z_{\lambda}^{-1}I -Z_{\lambda}^{-\lambda} H/\left\|H \right\| \right\}]\\
&\geq& Tr[Y^{1-\lambda}\left\{ \ln_{\lambda}Z_{\lambda}^{-1}I -Z_{\lambda}^{-\lambda} H/\left\|H \right\| \right\}]\\
&=& Tr[Y^{1-\lambda}\left\{ -H/\left\| H\right\| +\ln_{\lambda}Z_{\lambda}^{-1} \left(I-\lambda H/\left\|H\right\|\right) \right\}]\\
&=& Tr[Y^{1-\lambda}\ln_{\lambda}\left\{Z_{\lambda}^{-1}\exp_{\lambda}\left(-H/\left\|H\right\|\right) \right\}]\\
&=& Tr[Y^{1-\lambda}\ln_{\lambda}Y].
\end{eqnarray*}
By Eq.(\ref{nonnega_rela})
we have 
$$
S_{\lambda}(X) = -Tr[X^{1-\lambda}\ln_{\lambda}X] \leq -Tr[X^{1-\lambda}\ln_{\lambda}Y] \leq -Tr[Y^{1-\lambda}\ln_{\lambda}Y] = S_{\lambda}(Y).
$$
\hfill \qed

\begin{Rem}
Since $-x^{1-\lambda} \ln_{\lambda}x$ is a strictly concave function, $S_{\lambda}$ is a strictly concave function on the set $C_{\lambda}$.
This means that the maximizing $Y$ is uniquely determined so that we may regard $Y$ as a generalized Gibbs state,
since an original Gibbs state $e^{-\beta H}/Tr[e^{-\beta H}]$, where $\beta \equiv 1/T$ and $T$ represents a physical temperature, gives the maximum value of the von Neumann entropy.
Thus, we may define a generalized Helmholtz free energy by
$$
F_{\lambda}(X,H) \equiv Tr[X^{1-\lambda}H] - \left\| H \right\| S_{\lambda}(X).
$$
This can be also represented by the Tsallis relative entropy such as
$$
F_{\lambda}(X,H) = \left\| H \right\| D_{\lambda}(X\vert Y) + \ln_{\lambda}Z_{\lambda}^{-1} Tr[X^{1-\lambda}(\left\|H\right\|-\lambda H)].
$$
\end{Rem}

The following corollary easily follows by taking the limit as $\lambda \to 0$.

\begin{Cor} {\bf (\cite{Thi,UO})}
Let $Y = Z_{0}^{-1} \exp \left(-H/{\left\| H\right\|} \right) $, where $Z_{0} \equiv Tr[\exp \left(-H/{\left\| H\right\|} \right) ]$,
for an $n\times n$ Hermitian matrix $H$.
\begin{itemize}
\item[(i)] If $X \in \widetilde{ C_{0} }$, then $S_{0}(X) \leq \log Z_{0}.$
\item[(ii)] If $X \in C_{0}$, then $S_{0}(X) \leq S_{0}(Y).$
\end{itemize}
\end{Cor}

\section{On some trace inequalities related to Tsallis relative entropy}

In this section, we consider an extension of the following inequality \cite{HP2}:
\begin{equation} \label{hiai_petz_source}
 Tr[X(\log X + \log Y)] \leq  \frac{1}{p} Tr[X\log X^{p/2}Y^pX^{p/2}]
\end{equation}
for non-negative matrices $X$ and $Y$, and $p > 0$.

For the proof of the following Theorem \ref{the4_1}, we use the following famous inequalities.

\begin{Lem} {\bf (\cite{HP2})}  \label{hp_lemma}
For any Hermitian matrices $A$ and $B$, $0\leq \lambda \leq 1$ and $p>0$, we have the inequality: 
$$
Tr\left[\left(e^{pA}\sharp_{\lambda}e^{pB}\right)^{1/p}\right] \leq Tr\left[ e^{\left(1-\lambda\right)A+\lambda B}\right],
$$
where the $\lambda$-geometric mean for positive matrices $A$ and $B$ is defined by
$$
A \sharp_{\lambda} B \equiv A^{1/2}\left(A^{-1/2}BA^{-1/2}\right)^{\lambda}A^{1/2}.
$$
\end{Lem}

\begin{Lem} {\bf (\cite{Gol,Tho})}  \label{gt_lemma}
For any Hermitian matrices $G$ and $H$, we have the Golden-Thompson inequality: 
$$
Tr\left[e^{G+H}\right]\leq Tr\left[e^G e^H\right].
$$
\end{Lem}

\begin{The} \label{the4_1}
 For positive matrices $X$ and $Y$, $p \geq 1$ and $0 < \lambda \leq 1$, we have
\begin{equation}  \label{further_ineq1}
D_{\lambda}(X\vert Y) \leq -Tr[X\ln_{\lambda}(X^{-p/2}Y^pX^{-p/2})^{1/p}].
\end{equation}
\end{The}

{\it Proof}:
 First of all, we note that we have the following inequality \cite{Araki}
\begin{equation}
Tr[(Y^{1/2}XY^{1/2})^{rp}] \geq Tr[(Y^{r/2}X^rY^{r/2})^p] \label{araki_corollary}
\end{equation}
for non-negative matrices $X$ and $Y$, and $0 \leq r \leq 1, p >0$.
Similarly to the proof of Theorem 2.2 in \cite{FYK}, 
inequality (\ref{further_ineq1}) easily follows by setting $A=\log X$ and $B=\log Y$ in Lemma \ref{hp_lemma} such that
\begin{eqnarray}
Tr[(X^p\sharp_{\lambda}Y^p)^{1/p}] &\leq& Tr[e^{\log X^{1-\lambda}+\log Y^{\lambda}}] \nonumber \\
&\leq& Tr[e^{\log X^{1-\lambda}}e^{\log Y^{\lambda}}] \nonumber \\
&=& Tr[X^{1-\lambda} Y^{\lambda}], \label{proof_the4_1_1}
\end{eqnarray}
by Lemma \ref{gt_lemma}. 
In addtion, we have
\begin{equation} \label{proof_the4_1_2}
Tr[X^rY^r] \leq Tr[(Y^{1/2}XY^{1/2})^r],\,\,\,(0\leq r\leq 1),
\end{equation}
taking $p=1$ of inequality (\ref{araki_corollary}).
By (\ref{proof_the4_1_1}) and (\ref{proof_the4_1_2}) we obtain:
$$
Tr[(X^p\sharp_{\lambda}Y^p)^{1/p}] = Tr\left[\left\{ X^{p/2} (X^{-p/2}Y^pX^{-p/2})^{\lambda}X^{p/2}\right\}^{1/p}\right] \geq 
Tr[X(X^{-p/2}Y^pX^{-p/2})^{\lambda /p}].
$$
Thus we have,
\begin{eqnarray*}
D_{\lambda}(X\vert Y) &=& \frac{Tr[X-X^{1-\lambda}Y^{\lambda}]}{\lambda} \\
&\leq& \frac{Tr[X-X(X^{-p/2}Y^pX^{-p/2})^{\lambda /p}]}{\lambda} \\
&=& -\frac{Tr[X\left\{ ((X^{-p/2}Y^pX^{-p/2})^{1/p})^{\lambda} -I \right\} ]}{\lambda} \\
&=& -Tr[X\ln_{\lambda} (X^{-p/2}Y^pX^{-p/2})^{1/p}].
\end{eqnarray*}

\hfill \qed

\begin{Rem}
For positive matrices $X$ and $Y$, $0 < p < 1$ and $0 < \lambda \leq 1$, the following inequality dose not hold in general:
\begin{equation}  
D_{\lambda}(X\vert Y) \leq -Tr[X\ln_{\lambda}(X^{-p/2}Y^pX^{-p/2})^{1/p}].  \label{further_ineq1_conjecture}
\end{equation}
Indeed, the inequality (\ref{further_ineq1_conjecture}) is equivalent to
\begin{equation} 
Tr[X(X^{-p/2}Y^pX^{-p/2})^{\lambda /p}] \leq Tr[X^{1-\lambda}Y^{\lambda}]. \label{further_ineq1_conjecture2}
\end{equation}
Then we have many counter-examples. If we set $p=0.3$, $\lambda = 0.9$ and $
X = \left( \begin{array}{l}
 10\,\,\,\,3 \\ 
 \,3\,\,\,\,\,9 \\ 
 \end{array} \right),Y = \left( \begin{array}{l}
 5\,\,\,\,\,4 \\ 
 4\,\,\,\,\,5 \\ 
 \end{array} \right),
$
then inequality (\ref{further_ineq1_conjecture2}) fails. (R.H.S. minus L.H.S. of (\ref{further_ineq1_conjecture2}) approximately becomes -0.00309808.)
Thus inequality (\ref{further_ineq1_conjecture}) is not true in general.
\end{Rem}

\begin{Cor} 
\begin{itemize}
\item[(i)] For positive matrices $X$ and $Y$, the trace inequality
$$
 D_{\lambda} (X \vert Y ) \leq - Tr[X\ln_{\lambda}(X^{-1/2}YX^{-1/2})]
$$
holds.
\item[(ii)]  For positive matrices $X$ and $Y$, and $p \geq 1$, we have inequality (\ref{hiai_petz_source}).
\end{itemize}
\end{Cor}
{\it Proof}:
\begin{itemize}
\item[(i)] Put $p=1$ in (1) of Theorem \ref{the4_1}.
\item[(ii)] Take the limit as $\lambda \to 0$.
\end{itemize}
\hfill \qed


\section*{Acknowledgement}
The authour would like to thank the reviewer for providing valuable comments to improve the manuscript.
The authour would like to thank Professor K.Yanagi and Professor K.Kuriyama for providing valuable comments and constant encouragement.
This work was supported by the Japanese Ministry of Education, Science, Sports and Culture, Grant-in-Aid for 
Encouragement of Young Scientists (B), 17740068.
This work was also partially supported by the Ministry of Education,
Science, Sports and Culture, Grant-in-Aid for Scientific Research (B), 18300003.

\end{document}